\documentstyle[prd,aps,preprint]{revtex}
\begin{document}
\draft

%
%

\preprint{Nisho-02/2} \title{Vortex Excitations in $\alpha$ 
Cluster Condensed Nuclei}
\author{Aiichi Iwazaki}
\address{Department of Physics, Nishogakusha University, Shonan Ohi Chiba
  277-8585,\ Japan.} \date{May 1, 2002} \maketitle
\begin{abstract}
We point out an intriguing possibility that 
there exist topological soliton-like excitations in nuclei with
$\alpha$ cluster condensation. The excitations are 
vortices in superfluid, a property possessed by
such Bose condensed nuclei. 
We show that the vortex-like states possess 
quantum numbers $J^{P}=n^{(-1)^n}$ and energies $\sim $ several MeV,  
measured 
from the ground state energies of the $n \alpha$ condensed states.
We point out actual candidates, 
for instance, $2^{+}(3\,\mbox{MeV})$ for $^{8}$Be, $3^{-}(9.6\,\mbox{MeV})$
for $^{12}$C, etc, although the topological nature 
as a vortex is obscure in such light nuclei.
\end{abstract}
\hspace*{0.3cm}
\pacs{PACS 03.75.Fi, 11.27.+d, 21.10.Dr, 21.10.Re, 21.65.+f \\ 
Topological Soliton,\,\,$\alpha$ Cluster Condensation,\,\,
Nuclear Matter
\hspace*{1cm}}
\tightenlines
Topological excitations\cite{topo} arise in many branches of physics
and are fascinating objects in nature.
Especially, vortex excitations are well known
in superfluid such as He liquid, superconductor,
or in quantum Hall liquid\cite{iwa}.
These excitations are characterized as topological
solitons in Bose condensed matter 
with infinite degree of freedom.
Namely, they are coherent states of bosons,
e.g. He, Cooper pairs of electrons
or bosonized electrons. 
Even in a finite system of atomic gas
involving the order of $10^5$ atoms,
such vortex excitations have been observed\cite{bec}.
Now, an interesting question arises;
whether or not 
vortex state may exist in
a smaller system whose number of dynamical degrees
of freedom is such as $10^{4}, 10^{3}, 10^{2},10$ or $3,4$. 
We expect that vortex excitations or vortex-like excitations
are still present even in such a small system although
the topological properties ( topological stability, 
vorticity, etc. ) of the excitations become 
obscure gradually as the number of degrees of freedom 
in the system decreases. But some 
specific properties characterizing the vortex solitons
are preserved even in such states with small dynamical degrees of
freedom.
In this letter we point out that
vortex-like excitations may arise in nuclei with
$\alpha$ cluster condensation. 

It is well known that many nuclear states
involve cluster structures\cite{cluster} in themselves such as
$\alpha$, C, O etc. Among them,  
the states with the $\alpha$ cluster
condensation have recently been argued\cite{alpha} to exist
around threshold energies at which the nuclei are dissociated 
into each cluster. The excited states with larger radii 
than that of the ground state of $^{12}$C or 
$^{16}$O have been shown\cite{tohsaki} to   
involve possibly $\alpha$ clusters
interacting weakly with each other.
For example, a detail calculation has suggested that
the second $0^{+}$ state in $^{12}$C
is such a state with $ 3 \alpha$ clusters
with its rms radius $\simeq 4.3$ fm. Thus, the nuclear density 
is much smaller than that of the ground state whose radius
is about $2.5$ fm. The small density of the
nucleus implies that the $\alpha$ clusters interact weakly
with each other so that they may condense 
into an identical state. Similarly, the calculation 
have shown possible existence of $n \alpha$ clusters condensed nuclei
with larger mass numbers $4n$ ( $n\geq 3$ ).
  
Based on the possible existence of the $\alpha$ cluster 
condensed states in nuclei, we wish to see the vortex-like excitations
in the nuclei. Our concern is whether or not their energies
are accessible for observation in such small system. Furthermore,
we wish to clarify what quantum state do the clusters occupy in 
such a vortex-like nuclear state of the finite 
system, 
although topological excitations can only arise in infinite 
system, rigorously speaking. 

Topological states are characterized by 
nontrivial phases of the condensates and they are not  
eigenstates of the particle number. The number of the particles
in the states is fluctuating.
Nuclei are, however, the eigenstates of the number of nucleons. 
On the other hand, the number of clusters in some nuclear states
is not necessarily definite. The wave functions of the states involve 
non-cluster components as well as cluster components\cite{cluster}.
That is, the number of the clusters is fluctuating generally in such
nuclear states.
Therefore, it is reasonable to expect the presence of
such nuclear states involving
the vortex-like excitation.

Hereafter, we are only concerned with symmetric nuclei ( $N=Z$ ).
Some excited states of such nuclei are supposed to be $\alpha$ condensed states.
We use a model of a complex boson field $\phi$ in order to 
describe condensed states and vortex excitations.  
The boson field represents quanta of $\alpha$ clusters which
interacts with each other through a contact interaction in the model.
The Hamiltonian is supposed to be given by\cite{vortex}

\begin{equation}
H=\int d^3x\biggl(\frac{|\vec{\partial}\phi|^2}{2M}+\frac{\lambda(|\phi|^2-\rho)^2}{2}\biggr)
+V_c
\end{equation}
where $M$ ( $\rho$ ) denotes mass of $\alpha$ particle ( average
number density of $\alpha$ clusters in Bose condensed nuclei ). The term $|\phi|^4$
represents the contact interaction between $\alpha$ clusters, i.e.
$V(x-y)=\lambda \delta^3(x-y)$. The strength $\lambda$ can be
extracted from Gaussian potentials\cite{potential} and is given by 
$1.7\times 10^{-4}/(\mbox{MeV})^2$. The term $V_c$ represents
Coulomb interaction: $V_c=e^2\int (2|\phi (x)|^2)\,\frac{1}{2|x-y|}\,
(2|\phi (y)|^2)\, d^3x\,d^3y-3(2ne)^2/5R_0$ where the last term represents
the Coulomb energy of spherical nuclei  
with uniform charge distributions.  
We are only concerned with the difference
between the energy of the ground state with $\alpha$ condensation and 
the energies of the excited states with vortices.

We have assumed in the treatment of the $\alpha $ particles that the $\alpha$ particles interact very weakly in
the $\alpha$ condensed nuclei and that their internal structures
can be neglected; they are weakly interacting point particles.   
The assumption seems to be reasonable because the proposed $\alpha$ condensed nuclei 
have much smaller nuclear density\cite{tohsaki} ( $\sim 1/5\times
\mbox{standard nuclear density} $) so that the momentum of each cluster is
not so large as the internal structure of the cluster to affect seriously 
the properties of the condensed states.
Furthermore, we assume hereafter that the properties of the finite nuclei can be
extracted from those of infinite nuclear matter. Actually, the
Hamiltonian describes the infinite $\alpha$ clusters matter.
In order to describe
precisely the finite $\alpha$ condensed nuclei we should use more
realistic potentials involving attractive potentials.
In our model only repulsive potential is included and
this is sufficient for the formation of
the $\alpha$ condensation and  the vortex in the infinite nuclear matter.
Obviously,
attractive potentials for the formation of the $\alpha $ condensed
finite nuclei
are needed. Thus, more realistic treatment by which our
analysis is now in progress, is 
to use the Gaussian potentials in \cite{potential}.
Such a treatment leads to vortices as non-topological solitons.

We comment that Coleman's Q ball\cite{Q} is also non-topological
solitons. His model Hamiltonian involves also attractive potentials
of fields. Especially, they are point attractive interactions, while
ours are nonlocal potentials such as Gaussian potentials. 
As far as we know, such nonlocal potentials have not been
used for the formation of nontopological solitons.
But they are inevitably needed for the formation of such solitons
in finite nuclei.

The ground state of the Hamiltonian is given by 
$<\phi>\simeq \phi_0=e^{i\theta_0}\sqrt{\rho}$ with 
a constant $\theta_0$. 
Obviously, this represents an infinitely large
condensed state of the $\alpha$ clusters.    
In order to describe finite nuclei we simply  
cut the finite matter out of this infinite matter.  
Thus, the mass number $A$ of the nuclei under consideration
is given by $A=4n=\int d^3x |\phi_0|^2 =4\pi R_0^3 \rho/3$ where $R_0$ is
the radius of the $n \alpha$ condensed nucleus with no vortex
excitations. 
Here, the form of 
the nucleus is assumed to be spherical. 

Vortex excitation is characterized by non trivial phase factor
$e^{i\theta}$ in $\phi$ with angle $\theta$ around $z$ axis.
Namely, this factor induces a current $\sim \vec{\partial} \theta$ of $\alpha$ clusters around
the center of the vortex; the center is along $z$ axis.
In order to
find such vortex excitations, 
by taking $\phi=\sqrt{\rho}f(r)e^{i\theta}$, 
we solve an equation derived from the Hamiltonian,

\begin{equation}
\label{eq2}
  -\frac{1}{2M}\vec{\partial}^2\phi+\lambda (|\phi|^2-\rho)\phi
  =\rho e^{i\theta}\{-\frac{1}{2M}\biggl(f(r)^{\prime\prime}+\frac{f(r)^{\prime}}{r}
-\frac{1}{r^2}\biggr)+\lambda \rho(f(r)^2-1)f(r)\}=0,
\end{equation}   
where $r$ denotes radial coordinate in x-y plane and the prime denotes
a derivative in $r$. We have assumed axial symmetric solutions which
has no dependence on the coordinate $z$.
The Coulomb interaction is taken into account perturbatively;
after finding the solutions we calculate its Coulomb energy.

In order to avoid a singularity at $r=0$, $f(r)$
has to satisfy the boundary condition, $f(r=0)=0$. $f(r)$ should
also satisfy the boundary condition, $f(r=\infty)=1$. The latter
condition is, however, not necessarily needed for the finite nuclei although
the condition is needed for the infinite system in order to for  
the vortex energy per unit length being finite.  
It has to be
replaced by the condition $f(r=\infty)=0$ when we take into account a
realistic potential of the $\alpha$ clusters. But since the
spatial extension ( coherent length $=1/(\sqrt{2\lambda M \rho})\simeq 
0.4$\,fm ) 
of the vortex is much less than the radius of the 
$\alpha$ condensed nucleus $\sim 4$ fm,  
the condition $f(r=\infty)=1$ should be satisfied effectively.
Here, we adopt this condition for simplicity.
( In real finite system $f$ starts with $0$ at $r=0$ and goes to
$1$ as $r$ goes beyond the coherent length. Finally, 
it goes to zero as $r
\to \infty$. Thus in our approximation
 we neglect surface effects of the nuclei
although it may be 
important in small nuclei.) 
In our treatment the nucleus with the vortex excitation 
is supposed to be an axial symmetric ball; the density of $\alpha$
cluster does not depend on both $z$ and $\theta$, but depends on $r$
in the spherical ball. 
Although it is very
crude approximation, this simple treatment 
reveals significant properties of the states with vortices.     

The vortex solutions of the equation (\ref{eq2}) can be obtained numerically.
The energies of those solutions are obtained by inserting the solutions
into the Hamiltonian,

\begin{eqnarray}
E_v&=&\frac{2\pi\rho}{M\sqrt{M\lambda \rho}}K, \quad  K=\int_{0}^{S_v} ds
s\sqrt{S_v^2-s^2}\biggl((f^{\prime})^2+f^2/s^2+(f^2-1)^2\biggr) \\
 &\simeq&6.2\,\mbox{MeV}\, \sqrt{\frac{n}{3}}\,
\biggl(\frac{4\,\mbox{fm}}{R_0}\biggr)^{3/2}(S_v/7.1)(\log(S_v/7.1)+1)
\end{eqnarray} 
with $S_v=\sqrt{\lambda M \rho}\,R_v=3.8 \sqrt{n}(R_v/R_0)\sqrt{4\,\mbox{fm}/R_0}$,
where we have not included their Coulomb energies in this formula.
$R_v$ is the radius of the nucleus with the vortex and 
is defined such that $n=\int |\phi|^2 d^3x\simeq
\frac{0.05}{\sqrt{n}}(R_0/4\,\mbox{fm})^{3/2}\int^{S_v}_0 sds 
\sqrt{S_v^2-s^2}f[s]^2$.
$E_v$ represents the energy of the state with the vortex excitation
measured from the energy of $\alpha$ condensed state $<\phi>=\phi_0$
without the vortex.
We note that since the radius $R_v$ ( and $R_0$ ) may becomes large such as $R_v\sim
A^{1/3}$ because of
the constancy of the nuclear density even in the 
Bose condensed nuclei, the energy of the excited state with the vortex
goes such as $E_v\propto K \propto A^{1/3}=(4 n)^{1/3}\propto R_v$. Namely,
when the density $\rho$ does not depend on $n$,
$E_v$ approaches the energy of the vortex in the infinite system where
the vortex energy is proportional to its length $R_v$.

We do not still take account of the contribution of the Coulomb energy. But,
we find that the difference between the Coulomb energy of the vortex state
and the state $<\phi>=\phi_0$
is much less than $1$ MeV. This is because the radius $R_v$ of the vortex state 
is nearly equal to $R_0$ and so
the Coulomb energies of the both states 
are almost identical. $R_v$ 
becomes
slightly larger than $R_0$ owing to
the lack of the $\alpha$ clusters 
around the center of the vortex; $R_v/R_0\simeq 1.05$ 

Up to now, we use the terminology of ``nuclear state with vortex''.
Since the number of the alpha particles in nuclei discussed below is 
at most 4 or 5, topological nature is very obscure. But there are states
possessing  some properties of the vortex solitons. Thus,
we should use a terminology of ``vortex-like excitation''.

$^{12}$C has been argued\cite{tohsaki} to possess the condensed states of
3 $\alpha$ clusters. Since its radius is $\sim 4$ fm,
the state with the vortex-like excitation has the energy 
$\sim 6$ MeV measured from the energy of the condensed state.
Similarly, the state with the vortex in $^{16}$O has the energy
$\sim 9$ MeV. We expect in general that the energies of the states
with the vortex-like excitations increase 
such as $E_v \sim 6 (n/3)^{1/3}$ MeV with the number $n$ of $\alpha$ 
clusters, since $E_v \propto R_v \propto (4n)^{1/3}$.
It is important to stress that although our numerical results
come from very rough approximations, they indicate the energies of
such vortex-like states  sufficiently small for experimental detection 
( probably, 
several MeV $\sim 10 $ MeV for $^{12}$C, $^{16}$O,
or $^{20}$Ne smaller than the above estimation because attractive
interactions should be included in more realistic treatments 
).

We should mention that since these topological excitations are 
absolutely stable in
the infinite system, they may be rather 
stable even in the finite system. Namely, 
they ( $A=4 n$ ) would decay mainly into the lighter vortex-like
states ( $A=4 (n-1)$ ) 
by emitting an
$\alpha$ particle so as to preserve the vorticity, although the vorticity
is obscure in the finite system. 
If this decay mode
is inhibited energetically, 
the other decay
mode would be the decay with $\gamma$ emission. This corresponds to 
the disappearance of the vortex due to the finiteness
of the system;
the vortex gets away from the condensed state without passing 
infinite energy barrier.
This decay mode is relatively rare. Thus the decay width of the vortex-like nuclei
would be small when $\alpha $ emission is forbidden.  
This tendency can
be seen more prominently in heavier nuclei, since the number of
alpha particles increases and the vorticity becomes
clearer in such heavier nuclei.  

In order to see what quantum states each cluster occupies
in the vortex-like nuclei, we calculate
the angular momentum of the vortex soliton, especially, 
the $z$ component of the angular momentum. It is easy to find

\begin{equation}
J_z=\int d^3x (x P_y-y P_x)=\int d^3x |\phi|^2=n
\end{equation} 
with no use of the explicit form of the solution,
where the momentum $P_i$ of the field is given by
$P_i=\frac{1}{2i}(\phi^{\dagger} \partial_i \phi-\partial_i
  \phi^{\dagger}\, \phi)=|\phi|^2\partial_i \theta$. 
We also find $J_x=J_y=0$ with the use of the 
assumption of the axial symmetry of the solution.
The result implies that all clusters in the finite nucleus take an 
identical state with angular momentum $l=1$ and
rotate around $z$ axis, i.e. $l_z=+1$. We notice the configulation of the 
vortex field $\phi \propto e^{i\theta}$, which indicates that 
the states with $l_z=1$ compose the vortex-like state.
Therefore, we expect that
the states with the vortex-like excitation are such states with
$J^{P}=3^{-}$ for $^{12}$C, $J^{P}=4^{+}$ for $^{16}$O, etc.
Generally, $J^{P}=n^{(-1)^{n}}$ for the nucleus $A=4n$.
All of the clusters occupy a p state with $l_z=1$
so that their wave functions vanish at the center axis ( $r=0$ ) of
the nucleus. 
This lack of clusters at 
the center axis of the nucleus is consistent with the form of
the vortex solution, i.e. $f(r=0)=0$. In this sense
the states have the similar property to that of
the vortex soliton.

Now, we wish to find some candidates for the vortex-like nuclear
states. The states possess quantum number $J^{P}=n^{(-1)^n}$ with $J_z=n$ 
and energies $\sim $ several MeV measured from the $\alpha $ cluster
condensed nuclear states. First, we take $^8$Be for which 2 alpha
clusters play. The state $2^+(3\,\mbox{MeV})$ is a candidate for the
nuclear vortex-like state. The state is well known as a molecular state 
with large nuclear radius and decays into two alphas. This state 
is the lightest vortex-like nuclear state, although its character as a
vortex is very obscure. Second, the state $3^{-}(9.6\,\mbox{MeV})$
is a candidate for the vortex-like nuclear state of $^{12}$C. The state is
also well known theoretically as a molecular state of equilateral triangle.
Each cluster occupies a p state with $l_z=1$. This molecular state is not so rigid 
that each cluster can move easily from its original position
at the corner of the equilateral triangle. The molecular state is 
soft. Therefore, we may
regard the state as a vortex-like state. Furthermore, its main decay mode is 
the emission of gamma ray with its width being small.   
This fact also supports our interpretation.
Finally, we point out a candidate for the nuclear vortex-like state of $^{16}$O.
This is the state $4^+( 16\,\mbox{MeV})$, which is not yet fully
understood. The state will be understood as a soft molecular state
of equilateral square. This 
is a our prediction based on the idea of the nuclear vortex-like states.



Finally we mention that the vortex state with higher
vorticity $\phi \sim e^{im\theta}$ ( $m>1$ ) may exist
with much higher energy as well as the topological state
possessing two vortices with each vorticity of $e^{i\theta}$. 
These are states of $\alpha$ clusters
occupying higher angular momentum ( $l>1$ ). These vortex states
would have various decay modes satisfying 
the conservation of the vorticity. We also point out that
our formulation is based on hypothetical $\alpha$ cluster condensed
nuclei. Although they have been argued to arise around the threshold energy,
our results hold even if they arise in any regions of energies. 
It is difficult, in general, 
to confirm the existence of such Bose condensed nuclei. 
But the observation of the topological 
nuclear states above the condensed states result in the confirmation of the 
superfluidity of the Bose condensed nuclei.

To summarize, we found the following picture of the 
$\alpha$ cluster condensed finite nuclei with vortex-like excitations.
Their excitation energies are several MeV$\sim 10$ MeV
for $^{12}$C $\sim $ $^{20}$Na
measured from the $\alpha$ condensed states without vortices.  
Since all $\alpha$ clusters occupy an p state with $l_z=1$ 
in the vortex-like nuclei,  
the states are characterized by $J^{P}=n^{(-1)^{n}}$ for the nuclei
with the mass number $A=4n$. Their main decay modes
are expected such that a vortex-like state with $J^{P}=n^{(-1)^{n}}$ decays
predominantly into a vortex-like state with 
$J^{P}=(n-1)^{(-1)^{n-1}}$ with emission of an $\alpha$ particle. 
But the decay width is small 
because the $\alpha$ particle emitted is dominantly in a s state. 

We thank Y. Akaishi, O. Morimatsu and Y. Kanada-En'yo in KEK for fruitful discussions
and H. Nemura for useful information of $\alpha$ particles.
We also thank N. Itagaki for useful imformations of nuclear molecular 
states in $^{12}$C and $^{16}$O.



\begin{thebibliography}{99}
\bibitem{topo}R. Rajaraman. Soliton and Instantons ( North
  Holland. Amsterdam 1982 ).
\bibitem{iwa}S.M. Girvin. The Quantum Hall Effect: Novel Excitations 
and Broken Symmeties in Les Houches Summer School 1998 ( Springer
Verleg 1999 ).\\
F.Z. Ezawa M. Hotta and A. Iwazaki, Phys. Rev. B46, 7765. 
\bibitem{bec}M.R. Matthews, etal, Phys. Rev. Lett. 83, 2498 (1999).
\bibitem{cluster}Y. Fujiwara, H. Horiuchi, K. Ikeda, M. Kamimura, 
K. Kato, Y. Suzuki and E. Uegaki,  Prog. Theor. Phys. Supplement
No.68, 29 (1980).
\bibitem{alpha}G. Ropke, A. Schnell, P. Schuck and P. Nozieres,
Phys. Rev. Lett. 80, 3177 (1998); M. Beyer, S.A. Sofianos, C. Kuhrts,
G. Ropke and P. Schuck, Phys. Lett. B488, 247 (2000).
\bibitem{tohsaki}A. Tohsaki, H. Horiuchi, P. Schuck and G. Ropke,
  Phys. Rev. Lett. 87, 192501 (2001).
\bibitem{vortex}H.B. Nielsen and P. Olesen, Nucl. Phys. B61, 45
  (1973).
\bibitem{potential}S. Ali and A.R. Bodmer, Nucl. Phys. 80, 99 (1966).
\bibitem{Q}S. Coleman, Nucl. Phys. B262, 263 (1985); Nucl. Phys. B269,
  744 (1986).
\end{thebibliography}
\end{document}